\documentclass[%
reprint,
 amsmath,amssymb,
 aip,
]{revtex4-2}

\usepackage{graphicx}
\usepackage{dcolumn}
\usepackage{bm}
\usepackage{glossaries}
\usepackage{booktabs}
\usepackage{nicefrac}
\usepackage{listings}

\newacronym{sm}{SM}{supplementary material}
\newacronym{paw}{PAW}{projector augmented wave}
\newacronym{bz}{BZ}{Brillouin zone}
\newacronym{tdl}{TDL}{thermodynamic limit}
\newacronym{cc}{CC}{coupled cluster}
\newacronym{hf}{HF}{Hartree-Fock}
\newacronym{dmc}{DMC}{Diffusion Monte Carlo}
\newacronym{ppl}{ppl}{particle-particle ladder}
\newacronym{rpa}{RPA}{random phase approximation}
\newacronym{ccsd}{CCSD}{coupled cluster with single and double particle–hole excitation operators}
\newacronym{ccsdpt}{CCSD(T)}{coupled cluster with single, double and perturbative triple particle–hole excitation operators}
\newacronym{bvk}{BvK}{Born-von Karman}
\newacronym{bla}{BLA}{bond length alternation}

\newcommand{\D}{\text d}
\newcommand{\ImI}{\text i}
\newcommand{\EuE}{\text e}

\begin{document}

\title{Sampling the reciprocal Coulomb potential in finite and anisotropic cells}

\author{Tobias Sch\"afer}
\email{tobias.schaefer@tuwien.ac.at}
\affiliation{Institute for Theoretical Physics, TU Wien, Wiedner Hauptstraße 8-10/136, A-1040 Vienna, Austria}

\author{William Z. Van Benschoten}
\affiliation{Department of Chemistry, University of Iowa, Iowa City, Iowa 52242, United States}

\author{James J. Shepherd}
\affiliation{Department of Chemistry, University of Iowa, Iowa City, Iowa 52242, United States}

\author{Andreas Gr\"uneis}
\affiliation{Institute for Theoretical Physics, TU Wien, Wiedner Hauptstraße 8-10/136, A-1040 Vienna, Austria}

\begin{abstract}
We present a robust strategy to numerically sample the Coulomb potential in reciprocal space for periodic Born-von Karman cells of general shape.
Our approach tackles two common issues of plane-wave based implementations of Coulomb integrals under periodic boundary conditions, the treatment of the singularity at the Brillouin-zone center, as well as quadrature errors, which can cause severe convergence problems in anisotropic cells, necessary for the calculation of low-dimensional systems.
We apply our strategy to the \gls{hf} and \gls{cc} theory and discuss the consequences of different sampling strategies on the different theories. 
We show that sampling the Coulomb potential via the widely used probe-charge Ewald method is unsuitable for \gls{cc} calculations in anisotropic cells.
To demonstrate the applicability of our developed approach, we study two representative, low-dimensional use cases: the infinite carbon chain, for which we report the first periodic CCSD(T) potential energy surface, as well as a surface slab of lithium hydride, for which we demonstrate the impact of different sampling strategies for calculating surface energies.
We find that our Coulomb sampling strategy serves as a vital solution, addressing the critical need for improved accuracy in plane-wave based CC calculations for low-dimensional systems.

\end{abstract}

\maketitle

\section{Introduction}

Coulomb integrals manifest in numerous theories and models within the domains of many-body physics and quantum chemistry. 
However, under periodic boundary conditions, numerically modelling the periodic Coulomb potential is a non-trivial endeavour.
While its well-known reciprocal form of $4\pi/q^2$ is beyond question, there is no ideal strategy to sample this function in convolutions with electron densities on a discretized grid.
Such samplings are necessary for the implementation of computational methods such as the \gls{hf} theory, M{\o}ller-Plesset perturbation theory \cite{Moeller1934} or \gls{cc} theories \cite{Coester1960,Cizek1966,Shavitt2009}, to name only a few.
While the integrable singularity (at $q=0$) is the subject of several previous investigations \cite{Gygi1986,Martyna1999,Paier2005,Carrier2007,Spencer2008,Broqvist2009,Sundararaman2013,McClain2017}, another less studied problem arises when considering anisotropic \gls{bvk} cells  \cite{Xing2022,Xing2023}, as in the study of low-dimensional systems.
The main target of this work is to enable \emph{ab-initio} many-electron calculations for ground states of low-dimensional systems.

Here, we present an effective sampling technique for the reciprocal Coulomb potential, which is designed for use with anisotropic unit cells. We show how this alleviates the problems that can arise when running many-electron calculations for these systems. The modified potential is universal in character and can just as well be applied to cells of general shape.


\section{Theory} \label{sec:theory}

To illustrate the inherent difficulties of the periodic Coulomb potential, we choose the exact exchange energy expression of the Hartree-Fock theory and focus on the reciprocal space formulation, as it is common in plane-wave based implementations.
In the \gls{tdl} using atomic units, the exchange energy per unit cell, $V$, of a periodic system reads\cite{Sundararaman2013}
\begin{align}
E_x =& - \frac{1}{2} \sum_{ij}^\text{occ.} \int_{\Omega_\text{BZ}} \frac{\D^3 k}{\Omega_\text{BZ}}  \int_{\Omega_\text{BZ}} \frac{\D^3 q}{\Omega_\text{BZ}}  \int_{\mathbb R^3} \D^3 r_1 \int_V \D^3 r_2  \nonumber\\
&\times\frac{\varphi^*_{i\bm k}(\bm r_1) \varphi_{j\bm k-\bm q}(\bm r_1) \; \varphi^*_{j\bm k-\bm q}(\bm r_2)  \varphi_{i\bm k}(\bm r_2)  }{|\bm r_1 - \bm r_2|} \;. \label{eq:int}
\end{align}
This expression involves a summation over the occupied orbitals $\varphi_i, \varphi_j$, which are normalized in the unit cell, as well as two integrals over the first Brillouin zone with the volume $\Omega_\text{BZ}$. $\bm q$ is a momentum transfer vector.
The equivalent formulation in reciprocal space allows to replace the two continuous real-space integrals by one discrete sum over all reciprocal lattice vectors, $\bm G$,
\begin{align}
E_x =& - \frac{V}{2} \sum_{ij}^\text{occ.} \int_{\Omega_\text{BZ}} \frac{\D^3 k}{\Omega_\text{BZ}}  \int_{\Omega_\text{BZ}} \frac{\D^3 q}{\Omega_\text{BZ}}  \sum_{\bm G}  \nonumber\\
&\times \frac{4\pi}{(\bm G+\bm q)^2} \; \left| \rho_{ij}^{\bm k\bm q}(\bm G) \right|^2  \label{eq:eexg}\;,
\end{align}
where we introduced the Fourier transformed co-densities
\begin{equation}
\rho_{ij}^{\bm k\bm q}(\bm G) = \int_{V}\D^3 r \; \varphi^*_{i\bm k}(\bm r) \varphi_{j\bm k-\bm q}(\bm r)\; \EuE^{\ImI (\bm G+\bm q) \bm r} \;.
\end{equation}

In order to evaluate the exchange energy expression numerically, it is necessary to discretize the two Brillouin zone integrals. 
In this context, two prominent challenges manifest themselves.
First, a discretization inevitably introduces a quadrature error, and second, the presence of the notorious, albeit integrable singularity at $\bm G +\bm q = 0$ necessitates careful handling, if the co-densities contain a monopole. In the exchange energy expression this always happens for $i=j$, as $\rho_{ii}^{\bm k,\bm q=0}(\bm G=0) = 1$.
We note, that these issues originate solely from the used cell shape and discretization grids, which means that no specific class of material is affected, rather all calculations in which Coulomb integrals are used are affected.
While an extensive body of literature addresses the singularity issue \cite{Gygi1986,Martyna1999,Paier2005,Carrier2007,Spencer2008,Broqvist2009,Sundararaman2013,McClain2017}, less attention has been devoted to mitigating the quadrature error \cite{Xing2022,Xing2023}, especially in cases involving anisotropic cell shapes.

For each $i,j$, $\bm k$ and $\bm G$ one has to discretize the integral over $\bm q$ in the expression of the exchange energy in Eq. (\ref{eq:eexg}).
This can be re-cast as,
\begin{equation}
I := \frac{1}{\Omega_\text{BZ}} \int_{\Omega_\text{BZ}}\D^3 q \; v(\bm q) \, f(\bm q) \;,  \label{eq:I}
\end{equation}
which can be considered as the average value of the product of the two functions, $v(\bm q) := 4\pi / (\bm G+\bm q)^2$ and $f(\bm q) :=  \left| \rho_{ij}^{\bm k\bm q}(\bm G) \right|^2 $. 
A mesh of $N$ q-points, $\{\bm q_n\}$, is commonly introduced, in order to discretize Eq. (\ref{eq:I}) as,
\begin{equation}
I \approx \frac 1 N \sum_{n=1}^N   \; v(\bm q_n) \, f(\bm q_n) \;. \label{eq:Isum1}
\end{equation}
Typically, so called Monkhorst-Pack meshes \cite{Monkhorst1976} or uniform $\bm \Gamma$-centered meshes are considered, where $\bm \Gamma$ is the center of the \gls{bz}.
However, it should be noted that $v(\bm q)$ can vary significantly stronger between the q-points than the co-densities in $f(\bm q)$, in particular if $|\bm G|$ is small or if the mesh is not equally dense in all spatial dimensions (e.g. in anisotropic cells). 
This is only the case for the integral over the so called momentum transfer $\bm q$, not for the $\bm k$ integral. 
The following \emph{exact} reformulation of Eq. (\ref{eq:I}), where we simply split the integration range into $N$ distinct sub-spaces $\Omega_\text{BZ} = \cup_n^N \Omega^n_\text{BZ}$, suggests an improved strategy to discretize the $\bm q$ integral: 
\begin{equation}
I = \frac{1}{\Omega_\text{BZ}} \sum_{n=1}^N \int_{\Omega^n_\text{BZ}}\D^3 q \; v(\bm q) \, f(\bm q) \;.
\end{equation}
If we define the sub-spaces $\Omega^n_\text{BZ}$ as the volume around the points $\bm q_n$ of the previously introduced q-mesh, and assume $f(\bm q)$ to be roughly constant within the subspace $\Omega^n_\text{BZ}$, hence $f(\bm q)\approx f(\bm q_n)$, we can approximate
\begin{align}
I &\approx \frac{1}{\Omega_\text{BZ}} \sum_{n=1}^N f(\bm q_n) \; \int_{\Omega^n_\text{BZ}}\D^3 q \; v(\bm q)  \nonumber\\
&= \frac{1}{N} \sum_{n=1}^N f(\bm q_n) \, \bar v_n \label{eq:Isum2}
\end{align}
where we introduced $\bar v_n$, which only depend on $n$ and $\bm G$, as a \emph{mean} Coulomb potential of the subspace $\Omega^n_\text{BZ}$,
\begin{equation}
\bar v_n(\bm G) = \frac{1}{\Omega^n_\text{BZ}} \int_{\Omega^n_\text{BZ}}\D^3 q \;  \frac{4\pi}{(\bm G+\bm q)^2} \;. \label{eq:vn}
\end{equation}
This formulation, which we abbreviate as ''\emph{mean}'' in this manuscript, decouples the integration meshes for $f(\bm q)$ and $v(\bm q)$. 
While conventional q-meshes can be used to evaluate Eq. (\ref{eq:Isum2}), more sophisticated numerical integration techniques have to be applied to evaluate the deceptively simple integral in Eq. (\ref{eq:vn}).
In Sec. \ref{sec:impl} we describe our refined numerical integration technique.
Since $\bar v_n(\bm G)$ depends only on $n$ and $\bm G$, it can be pre-calculated and stored in memory once the lattice and the q-mesh is defined.
Using the \emph{mean} strategy from Eq. (\ref{eq:Isum2}), the final expression for the exchange energy for a mesh of $N$ k-points and q-points finally reads,
\begin{equation}
E_x = - \frac{V}{2} \frac{1}{N^2} \sum_{ij}^\text{occ.} \sum_{n,m=1}^N  \sum_{\bm G} \bar v_n(\bm G) \left| \rho_{ij}^{\bm k_m\bm q_n}(\bm G) \right|^2  \;.
\end{equation}
It is worth noting that this approach addresses both issues: the singularity, and the quadrature errors, which are dominated by the strongly varying Coulomb potential.

The \emph{mean} strategy can readily be applied to Coulomb integrals of any post-\gls{hf} method.
Here, we apply it to the Coulomb integrals for the equations of \gls{ccsdpt}. 
For more details on the implemented equations in \gls{ccsdpt} theory we refer to Ref. \onlinecite{Bartlett2007}.



We compare our \emph{mean} strategy with two other common strategies, which are summarized in Tab. \ref{tab:potentials}.
With ''\emph{probe}'' we abbreviate the so called probe-charge Ewald method \cite{Paier2005}.
In the \emph{probe} strategy a usual discretization is performed, as formulated in Eq. (\ref{eq:Isum1}), but for the singularity at $\bm G+\bm q = 0$ the difference between an analytically calculated self-energy of a \emph{probe}-charge and the corresponding numerically computed self-energy using the given q-grid is employed,
\begin{multline}
v_0 = \frac 1 {\Omega_\text{BZ}} \int_{\mathbb R^3} \D^3 q \, \frac{4\pi}{q^2} \varrho(\bm q) \\
- \frac 1 N \sum_{n=1}^N\sum_{\bm G\neq0} \frac{4\pi}{(\bm G + \bm q_n)^2} \varrho(\bm G + \bm q_n)\;, \label{eq:FSG}
\end{multline}
with the probe-charge density $\varrho(\bm q) = 2\sqrt{\alpha/\pi}\,\exp(-\alpha q^2)$.
Commonly, a strongly localized probe-charge is considered, i.e. $\alpha>0$ but very small, such that $v_0$ is almost independent of $\alpha$.
The \emph{probe} strategy is the default setting in the {\sc{vasp}} code and provides numerically comparable results as the auxiliary function strategy by Gygi and Baldereschi \cite{Gygi1986,Broqvist2009}.

Secondly, with "\emph{disr}" we refer to the strategy to simply neglect (``disregard``) the grid point at the singularity at $\bm G+\bm q = 0$ in the discretization (\ref{eq:Isum1}). 
The \emph{disr} strategy is the easiest way to achieve a functional periodic implementation of the Coulomb integral, and is used in some works \cite{McClain2017,Taheridehkordi2023}.

We do not consider the spherical truncation scheme \cite{Spencer2008}, since it is inadequate for anisotropic \gls{bvk} cells.
The Wigner-Seitz truncation strategy \cite{Sundararaman2013} is also not considered, as no implementation was available for this work.

While preparing this manuscript, it was brought to our attention that an analogous approach was used for GW calculations of excited states in Ref. \onlinecite{Marini2009}. While it would be of interest to make comparison between the method presented here and Ref. \onlinecite{Marini2009}, currently no technical capability exists to do so because the method developed here focuses on \gls{cc} ground-state calculations.

\begin{table}
    \centering
    \begin{tabular}{c|c}
        strategy & Coulomb potential\\
        \hline
        \emph{mean}  & 
            \parbox{4cm}{\begin{equation*}
                \frac{1}{\Omega^n_\text{BZ}} \int_{\Omega^n_\text{BZ}}\D^3 q \;  \frac{4\pi}{(\bm G+\bm q)^2} 
            \end{equation*}}\\\hline
        \emph{probe} & 
            \parbox{3cm}{\begin{equation*}
              \begin{cases}
                  \frac{4\pi}{(\bm G+\bm q_n)^2}  & \bm G+\bm q_n \neq 0\\
                  v_0 & \bm G+\bm q_n = 0 
               \end{cases}
           \end{equation*}}\\\hline
        \emph{disr}  & 
            \parbox{3cm}{\begin{equation*}
              \begin{cases}
                  \frac{4\pi}{(\bm G+\bm q_n)^2}  & \bm G+\bm q_n \neq 0\\
                  0 & \bm G+\bm q_n = 0 
               \end{cases}
           \end{equation*}}\\
    \end{tabular}
    \caption{Comparison of the considered Coulomb kernels for a given sampling grid $\{\bm q_n\}$ of the Brillouine zone. Note that $v_0$ was defined in Eq. (\ref{eq:FSG}) and $\Omega^n_\text{BZ}$ was defined in the text before Eq. (\ref{eq:Isum2}).}
    \label{tab:potentials}
\end{table}
\section{Methods and Implementation}
\subsection{Software and methods}

We implemented the \emph{mean} Coulomb potential strategy in the Vienna ab initio simulation package ({\sc{vasp}}) \cite{Kresse1996} and present the implementation in section \ref{sec:impl}.
We refer to the associated \texttt{INCAR} tag in {\sc{vasp}} as \texttt{LHFMEANPOT = T}.


Through the interface between {\sc{vasp}} and the {\sc{cc4s}} code \cite{cc4s}, we can directly compare \emph{mean}, \emph{probe}, and \emph{disr} in periodic \gls{hf} and \gls{cc} calculations. 
All calculations were performed using a plane-wave cutoff of $500\,\text{eV}$.
The following \gls{paw} pseudopotentials with frozen-core potentials ($\texttt{POTCAR}$ files) were employed for the test systems: $\texttt{PAW\_PBE C\_GW\_new}$, $\texttt{Li\_GW}$, and $\texttt{H\_GW}$.
The unoccupied space for the \gls{cc} calculations were compressed to $15$ / $12$ natural orbitals per occupied orbitals for the carbon chain / LiH system.
Natural orbitals are obtained on one-particle reduced density matrices obtained from the RPA \cite{Ramberger2019} or approximate MP2 level \cite{Gruneis2011a} of theory.
For the carbon chain, a highly effective basis-set correction scheme \cite{Irmler2021} was employed to approach the complete-basis set limit of the \gls{cc} correlation energies.
Using the \emph{mean} potential, we employed \gls{bvk} cells with a vacuum of $8 \,\text\AA$ and 20 / 10 replicas of the two-atomic unit cell to calculate the potential energy surface of the carbon chain at the \gls{hf} / CCSD correlation level.
The used \gls{bvk} cells for the LiH slab are discussed in the results section.

\subsection{Implementation of the mean potential strategy} \label{sec:impl}

A numerically highly accurate integration of the \emph{mean} potential of Eq. (\ref{eq:vn}) is described in the following, addressing the perils of its seemingly simple form.
We partition the integral into a term designed for a radial grid and a term designed for a uniform grid by means of the identity, $1 = \EuE^{-\gamma \bm q^2} + (1-\EuE^{-\gamma \bm q^2})$, so that we can write,
\begin{multline}
\frac{1}{\Omega^n_\text{BZ}} \int_{\Omega^n_\text{BZ}}\D^3 q \;  \frac{4\pi}{(\bm G+\bm q)^2} =\\
\frac{4\pi}{\Omega^n_\text{BZ}} \Bigg[ \int_0^{2\pi} \D\varphi \int_0^\pi\D\theta \int_0^\infty\D q \;  \frac{\bm q^2 \sin\theta\, \EuE^{-\gamma \bm q^2}}{(\bm G+\bm q)^2} \, \Theta_n(q,\theta,\varphi) \\
+  \int_{\Omega^n_\text{BZ}}\D^3 q \;  \frac{1-\EuE^{-\gamma \bm q^2}}{(\bm G+\bm q)^2} \,\Bigg]
\label{eq:expandI}
\end{multline}
where $\Theta_n(q,\theta,\varphi)$ is 1 if the point $(q,\theta,\varphi)$ is in the integration volume $\Omega^n_\text{BZ}$, and 0 otherwise.

The first integral on the right hand side of Eq. (\ref{eq:expandI}) is numerically implemented on a radial grid, whereas the second integral is implemented on a uniform grid along the axes of the reciprocal vectors. 
Note that no singularity occurs in either integrands.
While the Jacobian determinant of the spherical coordinates cancel the singularity in the first integrand, the second integrand approaches the finite value $(1-\EuE^{-\gamma q^2})/q^2 \rightarrow \gamma$ for $q \rightarrow 0$.
In our implementation we set $\gamma = 8\ln(10) \cdot L_\text{max}^2/(2\pi)^2$, where $L_\text{max}$ is the largest edge length of the \gls{bvk} cell.
This choice ensures that the factor $\EuE^{-\gamma q^2}$ reaches virtually zero inside the neighboring integration regions, $\Omega^n_\text{BZ}$, of the region containing the origin $\bm G + \bm q = 0$.
For all other integration regions, the first integral is virtually zero, and its numerical evaluation can be dropped. 

The second integral, using the uniform grid, varies little even at the origin, because of the numerator $1-\EuE^{-\gamma q^2}$ and by the choice of $\gamma$.
Thus a relatively small number of grid points can be used.
For the integration region containing the origin $\bm G+\bm q=0$, the number of grid points in direction $i$ is chosen according to $N_i \equiv 1000\,\text\AA / L_i$, where $L_i$ is the corresponding edge length of the \gls{bvk} cell in units of {\AA}ngstr{\"o}m.
Based on this number, we halve the number of grid points for each neighboring integration domain, $\Omega^n_\text{BZ}$, that moves away from the origin.

Finally, we note that smearing the step function $\Theta_n(q,\theta,\varphi)$ improves the convergence of the numerical integration on the radial grid significantly.
In our implementation, we multiply Fermi-Dirac-like functions, $\left( \exp(8d/\Delta q)+1 \right)^{-1}$, in each direction, to smear the sharp step of $\Theta_n$ at the boundaries of the integration regions $\Omega^n_\text{BZ}$.
Here $d$ represents the signed distance between an integration boundary in one direction to the grid point $(q,\theta,\varphi)$ and $\Delta q$ is the grid spacing of the $q$ coordinate.
We use the following radial grid in our implementation: $N_q = q_\text{max} / \Delta q \equiv 20 $ with $q_\text{max} = 2\pi/L_\text{max}$, and $N_\varphi = \lfloor  2\pi\cdot N_q \rceil = 126$, as well as $N_\theta = \lfloor\pi\cdot N_q\rceil = 63$.
The symbol $\lfloor ... \rceil$ denotes rounding to the next integer.

The described implementation of the \emph{mean} Coulomb potential strategy allows to achieve sub-meV accuracy of the Hartree-Fock energy per unit cell with respect to the number of grid points.
\section{Results and Discussion}

\begin{figure}
    \centering
    \includegraphics[width=1.0\linewidth]{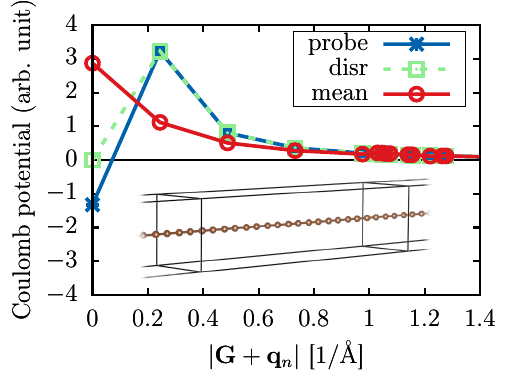}
    \caption{Comparison of the reciprocal Coulomb potential of the three different sampling strategies, \emph{mean}, \emph{disr}, and \emph{probe} (see text for abbreviations). A \gls{bvk} cell of the carbon chain with length $25.75\,\text\AA$ ($z$ dimension) and vacuum $6.0\,\text\AA$ ($x,y$ dimension) was used. Due to this geometry, the smallest reciprocal vectors have a magnitude of $|\bm G+\bm q_n| = 1.05\,\text\AA^{-1}$
    in $x,y$ direction and $0.24\,\text\AA^{-1}$ in $z$ direction.} 
    \label{fig:potentials}
\end{figure}

\subsection{Infinite carbon chain}\label{sec:chain}

We now turn to our first benchmark system, the infinite chain of carbon atoms in vacuum, where we aim for the calculation of the lattice constant and the \acrlong{bla}, which goes back to a Peierl's distortion \cite{Ramberger2021}.
To our knowledge, no canonical periodic \gls{ccsdpt} result was reported in the literature so far.
Based on an extrapolation of finite hydrogen-terminated chains, a \gls{ccsdpt} result of $0.125\,\text\AA$ was estimated for the infinite chain by Wanko et al. \cite{Wanko2016}.
However, this is in disagreement with the \gls{bla} of $0.136\,\text\AA$ and the lattice constant of $2.582\,\text\AA$ from a periodic \gls{dmc} calculation reported in Ref. \onlinecite{Mostaani2016}.
A further many-electron correlation calculation under periodic boundary conditions was reported by Ramberger and Kresse \cite{Ramberger2021} using the \gls{rpa}, resulting in a \gls{bla} of $0.129\,\text\AA$, but using a fixed lattice constant of $2.575\,\text\AA$.
Here, we report the first canonical and periodic CCSD(T) result for the \gls{bla} and the lattice constant.
We calculate five CCSD(T) energies each for \glspl{bla} in the range of $[0.075\,\text\AA,\,0.175\,\text\AA]$ and lattice constants in the range of $[2.425\,\text\AA,\,2.725\,\text\AA]$.
These 25 data points are then fitted using the function $E(b,a) = \sum_{n,m=0}^3 c_{nm} (b-b_0)^n (a-a_0)^m$, where $b$ and $a$ are the \gls{bla} and the lattice constant, respectively, and we explicitly neglect the coefficients $c_{11}=c_{31}=c_{32}=c_{13}=c_{23}=c_{33}=0$.
Optimizing the 12 fitting parameters, we find a \gls{bla} of $b_0=0.128\,\text\AA$ and a lattice constant of $a_0=2.578\,\text\AA$ with a standard deviation of the residuals of $0.5\,\text{meV}$ per unit cell.
The CCSD(T) data points and a heatmap of the potential energy surface can be found in the \gls{sm} \cite{sm}.



\begin{figure*}
    \centering
    \includegraphics[width=1.0\textwidth]{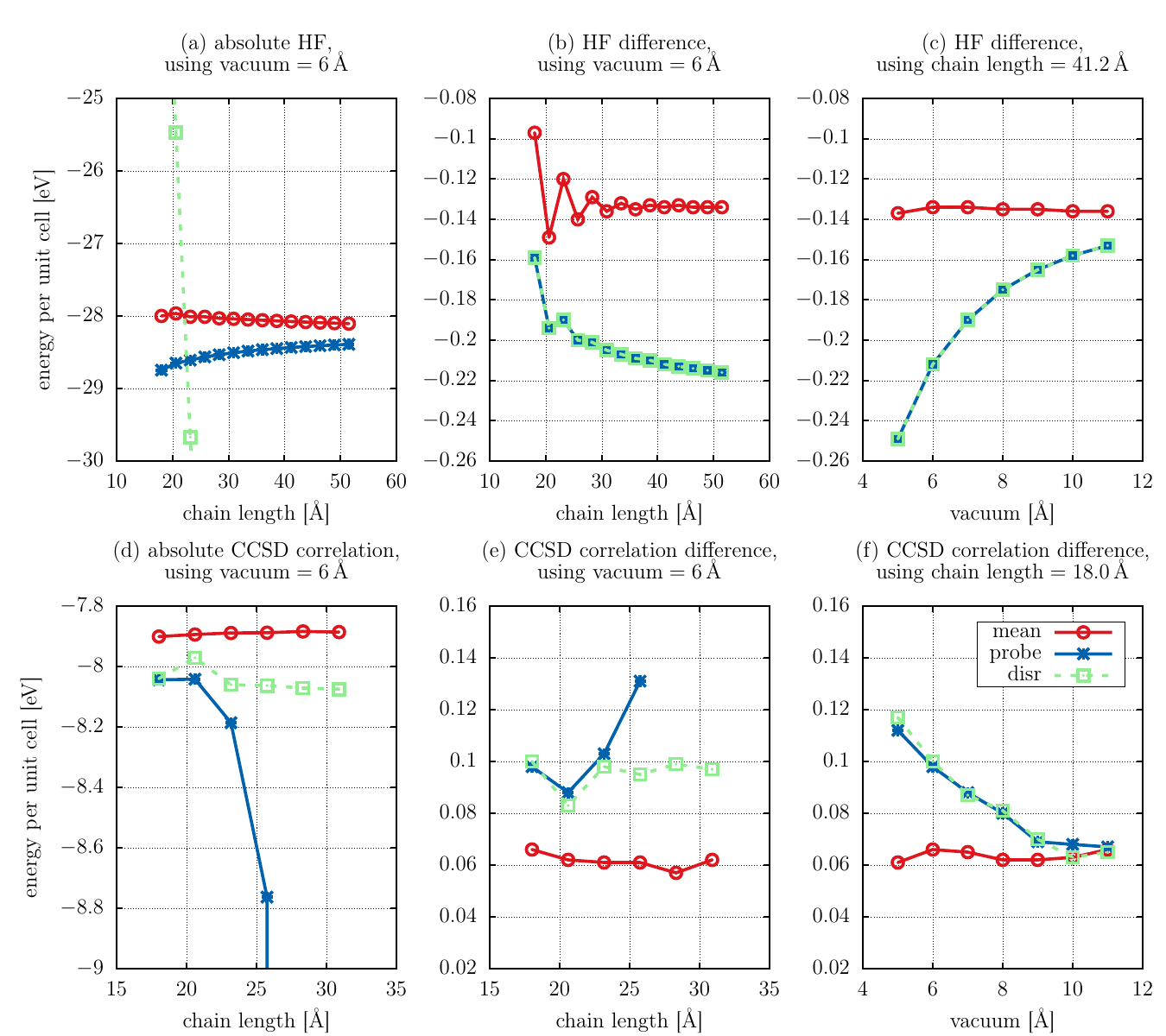}
    \caption{Convergence of the ground state energy of the carbon chain with respect to length (\gls{bvk} cell dimension in $z$ direction) and vacuum size (\gls{bvk} cell dimension in $x,y$ direction) using three different strategies to sample the periodic Coulomb kernel. Absolute energies refer to a \gls{bla} of $0.125\,\textrm\AA$, whereas differences refer to the difference between the \glspl{bla} $0.075\,\textrm\AA$ and $0.125\,\textrm\AA$. The graphs (a)-(c) show the convergence of the \gls{hf} contribution, wherease (d)-(f) show only the correlation contribution at the \gls{ccsd} level of theory.}
    \label{fig:carbyne}
\end{figure*}

These calculations require strongly anisotropic \gls{bvk} cells, which prohibited previous \gls{hf} and \gls{ccsdpt} calculations, the reasons for which will be evident below. 
For growing \gls{bvk} cells in the chain direction, $z$, the differences of the three strategies \emph{mean}, \emph{probe}, and \emph{disr} are evident for absolute HF energies, as can be observed in Fig. \ref{fig:carbyne}(a).
While \emph{disr} diverges, both \emph{mean} and \emph{probe} seem to converge to finite, though perhaps slightly different, limits.
This is not surprising as with fixed dimensions in $x$ and $y$, we are not approaching the \gls{tdl}, which should indeed be unique.

The divergence of \emph{disr} is due to the fact, that we approach the non-integrable singularity of $1/z^2$ at $z=0$. 
More precisely, this non-integrable singularity is approached if only the $z$ component of the \gls{bvk} cell is increased, turning only $\sum_z \rightarrow \int\D z$, while $\sum_{x,y}$ remains untouched.
Note that in three dimensions, $\D x\,\D y\,\D z / (x^2 + y^2 + z^2) = \D r \, \sin(\theta) \D \theta \, \D\varphi  $ is free from a singularity, while $\D z / (x^2 + y^2 + z^2)$ possesses a non-integrable singularity for the discrete sampling points $x=y=0$ at $z=0$.


The \emph{probe} strategy cancels this divergence of the non-integrable singularity and converges to a finite \gls{hf} result, as $v_0$ from Eq. (\ref{eq:FSG}) approaches $-\infty$ if only the $z$ dimension of the \gls{bvk} cell is increased.
Both, the value of $v_0$ and the \gls{hf} solution of the \emph{disr} strategy approach $-\infty$ for the same reason, as explained above.
The negative value of $v_0$ can be observed in Fig. \ref{fig:potentials}.
This \emph{probe} correction provides reasonable results at the \gls{hf} level of theory, but fails miserably at the \gls{ccsd} level, as can be seen in Fig. \ref{fig:carbyne}(d). 
Instead of converging to a limit, the \gls{ccsd} correlation energy falls into the negative and finally the solver fails to converge the \gls{ccsd} energy after 200 iterations, if the \emph{probe} strategy is used.
This is in contrast to \emph{mean} and \emph{disr}, for which the \gls{ccsd} solver successfully converges for each length of the \gls{bvk} cell.

We believe that negative values of $v_0$ are responsible for the deteriorated CCSD energies and for the failure of the \gls{ccsd} iterations.
Note, that the \emph{probe} strategy constructs effectively a pseudized Coulomb potential, as the correction is solely put into the $\bm G + \bm q=0$ component of the potential, here denoted as $v_0$.
While this pseudization manifests itself only as a shift in the total energy and in the orbital energies at \gls{hf} theory, it substantially changes the Coulomb integrals in the non-linear \gls{ccsd} equations.
This becomes striking for the Coulomb integrals in the so called ladder contributions.
A deeper analysis of the failure of the \gls{ccsd} iterations is beyond the scope of this work.
However, we emphasize that the divergence of $v_0$ into negative infinity, with increasing $z$ dimension of the \gls{bvk} cell, entails an underestimation of the mentioned Coulomb integrals.
It is conceivable that integrals become negative and thus represent an attractive rather than repulsive electron-electron interaction.

In contrast to the absolute \gls{hf} energies, the \gls{hf} energy differences converge only for the \emph{mean} strategy, as is apparent from Fig. \ref{fig:carbyne}(b).
Here an energy difference between the \glspl{bla} $0.075\,\text\AA$ and $0.125\,\text\AA$ is considered.
Furthermore, it becomes obvious, that both strategies \emph{probe} and \emph{disr} are equivalent for energy difference within the same \gls{bvk} cell at the \gls{hf} level.
This underlines, that $v_0$ simply introduces a constant energy shift in the \gls{hf} theory.

This is no longer true for the \gls{ccsd} theory, as evident from Fig. \ref{fig:carbyne}(e), which shows the \gls{ccsd} correlation contribution to this energy difference between the two \glspl{bla}.
While \emph{mean} provides relatively constant results, with mild but tolerable fluctuations, the failure of the \emph{probe} strategy is markedly visible.
Except for one strong outlier, the \emph{disr} strategy here appears to be only shifted to \emph{mean}.

We close the discussion of the convergence behaviour with the dependence on the size of the vacuum, i.e. the $x$ and $y$ dimension of the \gls{bvk} cell.
In Fig. \ref{fig:carbyne}(c), one can see that all strategies seemingly approach the same \gls{tdl} at the \gls{hf} level, although at significantly different rates.
Note, that the $z$ dimension of the \gls{bvk} cell is already quite large with $41.2\,\text\AA$.
This is also true at the \gls{ccsd} level of theory, as depicted in \ref{fig:carbyne}(f),
since we leave the strong anisotropic regime. Here, a reduced length of $18.0\,\text\AA$ was used.



\subsection{Lithium hydride surface}

\begin{figure}
    \centering
    \includegraphics[width=1.0\linewidth]{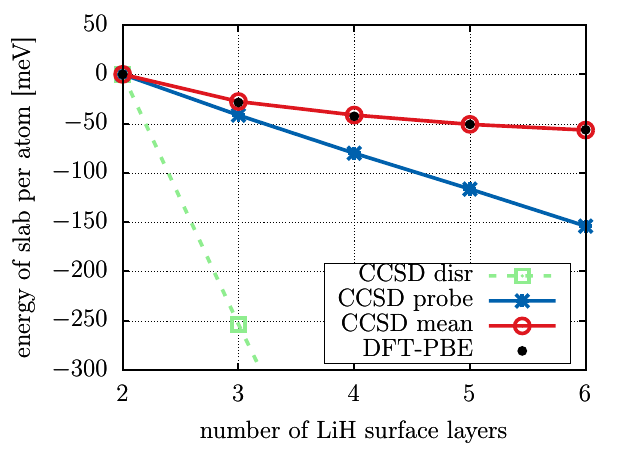}
    \caption{Convergence of the relative energy of a (001) LiH surface slab per atom with respect to the number of surface layers of the slab. The energies are shifted such that all values agree for the 2-layer slab.} 
    \label{fig:LiH}
\end{figure}

In our second benchmark system, we address the impact of the different sampling strategies for the electronic ground state of a surface slab model, as necessary for the calculation of surface formation energies or applications in heterogeneous catalysis. 
The electronic structure of a materials surface determines its physical and chemical properties, i.e. its functionality in scientific and industrial applications.
Under periodic boundary conditions, the \gls{bvk} cells used to model the surface slab can exhibit anisotropic shapes, primarily due to the necessary vacuum above the slab to prevent interactions between periodic images along the normal direction to the surface.
Here, we study the impact of the different sampling techniques \emph{mean}, \emph{probe}, and \emph{disr} on the convergence of the per-atom energy of a (001) lithium hydride (LiH) surface slab with respect to the number of surface layers.
This energy is a key ingredient for the calculation of surface free energies or for the simulation of surface reconstructions.
Note that we do not report a surface energy value for LiH in this study, as this requires calculations outside the primary concern of this work, including the determination of the bulk thermodynamic limit of LiH \cite{Mihm2021} and achieving the basis-set limit.

We model the LiH surface using a $2\times 2 \times n$ conventional slab, where we increase the number of layers $n$ from $2$ to $6$. 
Each layer consists of 16 atoms, and a vacuum of about $30\,\text\AA$ is added in the normal direction to the slab.
The \gls{bvk} cell of the $n=6$ layer slab is provided in the \gls{sm} \cite{sm}.

Figure \ref{fig:LiH} shows the relative CCSD energy per atom of the LiH surface slabs in dependence of the number of surface layers for different sampling techniques, as well as density function theory results using the Perdew-Burke-Ernzerhofer exchange-correlation functional (DFT-PBE) \cite{Perdew1996}.
For a better comparison of the convergence behaviour, all energies are shifted to match for the 2-layer slab.
For both \emph{probe} and \emph{disr} no convergence of the per-atom CCSD energy can be observed.
A differentiated view reveals, that in the case of \emph{probe} the \gls{hf} energy converges but not the \gls{ccsd} correlation contribution, and using \emph{disr} it is exactly the opposite.
This is documented in the \gls{sm} \cite{sm}.
In contrast, employing the \emph{mean} strategy, the CCSD results converges at nearly the same rate as DFT-PBE.
Note that periodic DFT-PBE implementations do not rely on two-electron four-orbital Coulomb integrals and therefore do not suffer from the discussed sampling issues. 
Thus, the agreement is a strong indication that the \emph{mean} strategy effectively eliminates quadrature errors.

\section{Summary and Conclusion} \label{sec:conclusion}

In this manuscript, we have introduced the \emph{mean} Coulomb potential as a sampling technique to enable many-electron ground-state calculations at the level of \acrlong{hf} and \acrlong{cc} theory in anisotropic \acrlong{bvk} cells.
By examining two illustrative cases, the infinite carbon chain and the (001) LiH surface, we identified and effectively resolved deficiencies inherent in existing sampling techniques for the reciprocal Coulomb potential.
The presented implementation for the evaluation of the \emph{mean} Coulomb potential is specifically designed for sub-meV accuracy of ground states and can be directly adopted in other codes employing the reciprocal form of the Coulomb potential.
This advancement not only enables predictive \emph{ab initio} calculations for low-dimensional systems using highly-accurate methods like \gls{ccsdpt}, but can be readily applied to \acrlong{bvk} cells of general shape.
Furthermore, the \emph{mean} potential technique can as well be applied without further ado to modified reciprocal Coulomb potentials, such as range-separated potentials \cite{Adamson1996,Paier2006} or truncated potentials \cite{Rozzi2006,Spencer2008,Sundararaman2013}, thereby both improving their sampling in reciprocal space and circumventing any special treatment of singularities.

\section*{Supplementary Material}

See the supplementary material \cite{sm} for the CCSD(T) data to calculate the potential energy surface of the carbon chain, a decomposition of the CCSD convergence of the (001) LiH surface slab energies into the \gls{hf} and correlation contribution, as well as the atomic structure of the 6-layer surface slab (POSCAR files).

\section*{Acknowledgements}

T.S. acknowledges support by the Austrian Science Fund through the project ESP 335-N.
The computational results presented have been largely achieved using the Vienna Scientific Cluster (VSC). 
The research presented here was funded in part by the National Science Foundation under NSF CHE-2045046 (J.J.S., W.Z.V).

\section*{Author declarations}

\subsection*{Conflict of Interest}

The authors declare no conflicts of interest.



\bibliography{refs}

\newpage\null\thispagestyle{empty}\newpage

\appendix

\section{CCSD(T) data of the carbon chain}

\begin{figure}
    \centering
    \includegraphics[width=1.0\linewidth]{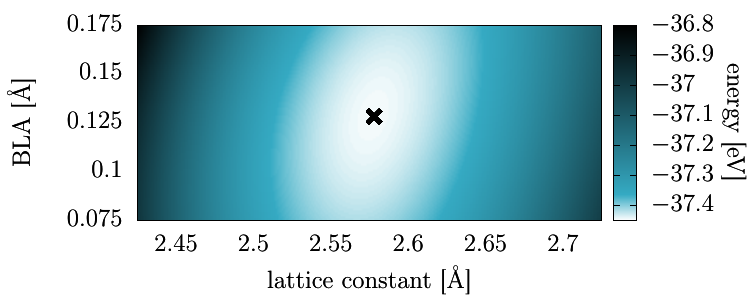}
    \caption{Heatmap of the potential energy surface of the infinite carbon chain per unit cell in dependence of the \gls{bla} and the lattice constant, calculated at the level of CCSD(T). The energy minimum at $2.578\,\text\AA$, $0.128\,\text\AA$ is denoted by a black x. The necessary information to reconstruct this potential energy surface are provided in the text.}
    \label{fig:carbyneheatmap}
\end{figure}

The CCSD(T) results of the carbon chain in dependence of the lattice constant and the \gls{bla} can be found in Table \ref{tab:carbynedata}.
This data can be used in order to reconstruct the potential energy surface depicted in Fig. \ref{fig:carbyneheatmap}.

\begin{table}
\centering
\begin{tabular}{l|rrrrr}
          & \(0.075\) & \(0.100\) & \(0.125\) & \(0.150\) & \(0.175\) \\
\hline
\(2.425\) & \(-36.972\) & \(-36.965\) & \(-36.934\) & \(-36.878\) & \(-36.793\) \\
\(2.500\) & \(-37.322\) & \(-37.331\) & \(-37.322\) & \(-37.293\) & \(-37.242\) \\
\(2.575\) & \(-37.413\) &\( -37.435\) & \(-37.444\) & \(-37.437\) & \(-37.414\) \\
\(2.650\) & \(-37.292\) & \(-37.325\) & \(-37.347\) & \(-37.359\) & \(-37.357\) \\
\(2.725\) & \(-37.003\) & \(-37.043\) & \(-37.077\) & \(-37.103\) & \(-37.119\) 
\end{tabular}
\caption{CCSD(T) ground state energies of the carbon chain per unit cell. A row corresponds to a fixed lattice constant, while a column corresponds to a fixed \gls{bla}. The units of energies and lengths are eV and \AA, respectively.}
\label{tab:carbynedata}
\end{table}

\section{Lithium hydride (001) surface}

\begin{figure*}
    \centering
    \includegraphics[width=0.49\linewidth]{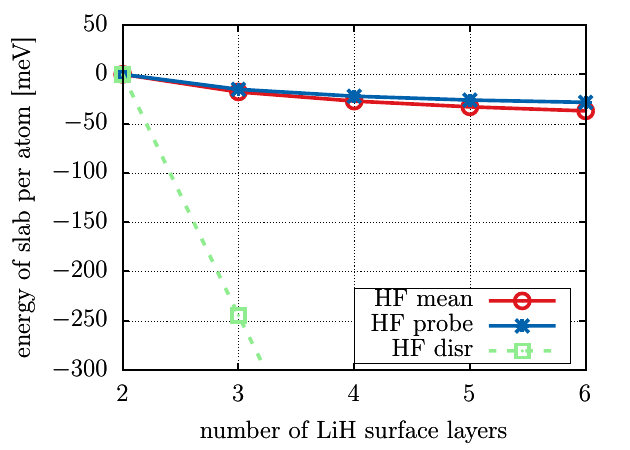}
    \includegraphics[width=0.49\linewidth]{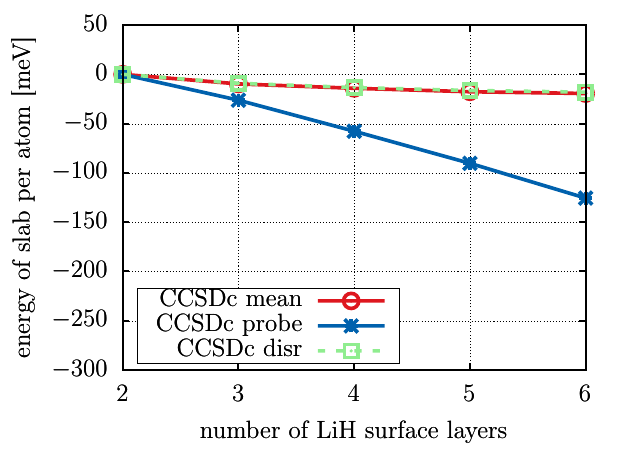}
    \caption{Convergence of the total energy of a (001) LiH surface slab per atom with respect to the number of surface layers of the slab. In the left plot only the HF contribution is shown, while the right plot shows only the CCSD correlation contribution. The total energies are shifted such that all values agree for the 2-layer slab.}
    \label{fig:surface}
\end{figure*}

The total HF and CCSD correlation energy per atom of the LiH surface slabs in dependence of the number of surface layer is shown in Fig. \ref{fig:surface}.
Only for \emph{mean} both the HF and the CCSD correlation contribution converge with respect to the number of surface layers.
In the case of the \emph{probe} strategy only the HF energy converges but not the CCSD correlation contribution.
In the case of the \emph{disr} strategy the HF energy diverges but the CCSD correlation contribution converges.

The structure of the 6-layer slab of LiH is provided below, in the form of a \texttt{POSCAR} file for {\sc{vasp}}:

\begin{lstlisting}
LiH
   1.00000000000000
     8.221600000  0.000000000  0.000000000
     0.000000000  8.221600000  0.000000000
     0.000000000  0.000000000 38.221600000
 Li H
 48 48
Cartesian
   0.00000000  0.00000000  0.00000000
   2.05540000  2.05540000  0.00000000
   2.05540000  0.00000000  2.05540000
   0.00000000  2.05540000  2.05540000
   0.00000000  0.00000000  4.11080000
   2.05540000  2.05540000  4.11080000
   2.05540000  0.00000000  6.16620000
   0.00000000  2.05540000  6.16620000
   0.00000000  0.00000000  8.22160000
   2.05540000  2.05540000  8.22160000
   2.05540000  0.00000000 10.27700000
   0.00000000  2.05540000 10.27700000
   4.11080000  0.00000000  0.00000000
   6.16620000  2.05540000  0.00000000
   6.16620000  0.00000000  2.05540000
   4.11080000  2.05540000  2.05540000
   4.11080000  0.00000000  4.11080000
   6.16620000  2.05540000  4.11080000
   6.16620000  0.00000000  6.16620000
   4.11080000  2.05540000  6.16620000
   4.11080000  0.00000000  8.22160000
   6.16620000  2.05540000  8.22160000
   6.16620000  0.00000000 10.27700000
   4.11080000  2.05540000 10.27700000
   0.00000000  4.11080000  0.00000000
   2.05540000  6.16620000  0.00000000
   2.05540000  4.11080000  2.05540000
   0.00000000  6.16620000  2.05540000
   0.00000000  4.11080000  4.11080000
   2.05540000  6.16620000  4.11080000
   2.05540000  4.11080000  6.16620000
   0.00000000  6.16620000  6.16620000
   0.00000000  4.11080000  8.22160000
   2.05540000  6.16620000  8.22160000
   2.05540000  4.11080000 10.27700000
   0.00000000  6.16620000 10.27700000
   4.11080000  4.11080000  0.00000000
   6.16620000  6.16620000  0.00000000
   6.16620000  4.11080000  2.05540000
   4.11080000  6.16620000  2.05540000
   4.11080000  4.11080000  4.11080000
   6.16620000  6.16620000  4.11080000
   6.16620000  4.11080000  6.16620000
   4.11080000  6.16620000  6.16620000
   4.11080000  4.11080000  8.22160000
   6.16620000  6.16620000  8.22160000
   6.16620000  4.11080000 10.27700000
   4.11080000  6.16620000 10.27700000
   0.00000000  2.05540000  0.00000000
   2.05540000  0.00000000  0.00000000
   0.00000000  0.00000000  2.05540000
   2.05540000  2.05540000  2.05540000
   0.00000000  2.05540000  4.11080000
   2.05540000  0.00000000  4.11080000
   0.00000000  0.00000000  6.16620000
   2.05540000  2.05540000  6.16620000
   0.00000000  2.05540000  8.22160000
   2.05540000  0.00000000  8.22160000
   0.00000000  0.00000000 10.27700000
   2.05540000  2.05540000 10.27700000
   4.11080000  2.05540000  0.00000000
   6.16620000  0.00000000  0.00000000
   4.11080000  0.00000000  2.05540000
   6.16620000  2.05540000  2.05540000
   4.11080000  2.05540000  4.11080000
   6.16620000  0.00000000  4.11080000
   4.11080000  0.00000000  6.16620000
   6.16620000  2.05540000  6.16620000
   4.11080000  2.05540000  8.22160000
   6.16620000  0.00000000  8.22160000
   4.11080000  0.00000000 10.27700000
   6.16620000  2.05540000 10.27700000
   0.00000000  6.16620000  0.00000000
   2.05540000  4.11080000  0.00000000
   0.00000000  4.11080000  2.05540000
   2.05540000  6.16620000  2.05540000
   0.00000000  6.16620000  4.11080000
   2.05540000  4.11080000  4.11080000
   0.00000000  4.11080000  6.16620000
   2.05540000  6.16620000  6.16620000
   0.00000000  6.16620000  8.22160000
   2.05540000  4.11080000  8.22160000
   0.00000000  4.11080000 10.27700000
   2.05540000  6.16620000 10.27700000
   4.11080000  6.16620000  0.00000000
   6.16620000  4.11080000  0.00000000
   4.11080000  4.11080000  2.05540000
   6.16620000  6.16620000  2.05540000
   4.11080000  6.16620000  4.11080000
   6.16620000  4.11080000  4.11080000
   4.11080000  4.11080000  6.16620000
   6.16620000  6.16620000  6.16620000
   4.11080000  6.16620000  8.22160000
   6.16620000  4.11080000  8.22160000
   4.11080000  4.11080000 10.27700000
   6.16620000  6.16620000 10.27700000
\end{lstlisting}

\end{document}